\documentclass[15pt a4paper]{article}

\usepackage{latexsym}
\usepackage{amsmath}
\usepackage{amssymb}
\usepackage{graphicx}
\graphicspath{{./}{./figs/}}

\begin{document}

\title{Projective synchronization in fractional order chaotic systems and its control\thanks{Progress of Theoretical Physics, Vol. 115, No. 3, March 2006, pp.661-666.}}
\author{Chunguang Li\thanks{Email: cgli@uestc.edu.cn}}

\date{\small Centre for Nonlinear and Complex Systems,
School of Electronic Engineering, \\University of Electronic
Science and Technology of China, Chengdu, 610054, P. R. China.}
\maketitle
\begin{abstract}
The chaotic dynamics of fractional (non-integer) order systems
have begun to attract much attention in recent years. In this
paper, we study the projective synchronization in two coupled
fractional order chaotic oscillators. It is shown that projective
synchronization can also exist in coupled fractional order chaotic
systems. A simple feedback control method for controlling the
scaling factor onto a desired value is also presented.

\end{abstract}

Although fractional calculus has a mathematical history nearly as
long as that of the integer-order calculus, the applications of it
to physics and engineering are just a recent focus of interest [1,
2]. Many systems are known to display fractional order dynamics,
such as viscoelastic systems [3-5], dielectric polarization [6],
electrode-electrolyte polarization [7] and electromagnetic waves
[8], so it is important to study the properties of fractional
order systems. The dynamics of fractional order systems has not
yet been fully studied, and it is by no mean trivial. The
definitions of fractional order calculus are geometrically and
physically less intuitive than the integer-order ones, and can't
be simulated directly in time-domain. It is still unclear whether
the dynamics of fractional order systems is similar to the
integer-order ones. More recently, many authors have begun to
investigate the chaotic dynamics of fractional order dynamical
systems [9-17]. In [9], it was shown that the fractional order
Chua's system of order as low as 2.7 can produce a chaotic
attractor. In [10], it was shown that nonautonomous Duffing
systems of order less than 2 can still behave in a chaotic manner.
In [11], chaotic behaviors of the fractional order ``jerk" model
was studied, in which chaotic attractor was obtained with system
orders as low as 2.1, and in [12] the control of this fractional
order chaotic system was reported. In [13], chaotic behavior of
the fractional order Lorenz system was studied, but unfortunately,
the results presented in this paper are not correct. In [14] and
[15], bifurcation and chaotic dynamics of the fractional order
cellular neural networks were studied. In [16], chaos and
hyperchaos in the fractional order R\"ossler equations were
studied, in which we showed that chaos can exist in the fractional
order R\"ossler equation with order as low as 2.4, and hyperchaos
can exist in the fractional order R\"ossler hyperchaos equation
with order as low as 3.8. In [17], we studied the chaotic behavior
and its control in the fractional order Chen system. In [18], the
author presents a broad review of existing models of fractional
kinetics and their connection to dynamical models, phase space
topology, and other characteristics of chaos.

On the other hand, synchronization of chaotic systems has
attracted much attentions [19] since the seminal paper by Pecora
and Carroll [20]. Besides the identical (complete) synchronization
of chaotic systems, some other types of synchronization are also
very interesting cooperative behaviors of chaotic systems, among
which are the phase synchronization [21], lag synchronization [22]
and projective synchronization [23] of chaotic oscillators.

In [24], we have studied the synchronization of fractional order
chaotic systems. And since then, the synchronization of fractional
order chaotic systems has begun to attract attentions of some
researchers, see for example [25]. But, in the literature, the
authors are all concerned with the identical synchronization of
fractional order chaotic systems. In [26], we studied the phase
and lag synchronization of coupled fractional order chaotic
oscillators. However, to our knowledge, projective synchronization
in fractional order chaotic systems has not been discussed yet. In
this paper, we address this topic.

Projective synchronization is the dynamical behavior in which the
response of two identical systems synchronize up to a constant
scaling factor. This phenomenon was observed in the coupled
integer-order partially linear systems
\begin{equation}
\begin{array}{l}
\dot{\mathbf{u}}_m=\mathbf{M}(z)\cdot{\mathbf{u}}_m,\\
\dot{z}=f(\mathbf{u}_m,z),\\
\dot{\mathbf{u}_s}=\mathbf{M}(z)\cdot\mathbf{u}_s
\end{array}
\end{equation}
where the matrix $\mathbf{M}(z)$ is only dependent on $z$,
$\mathbf{u}_m$ and $\mathbf{u}_s$ are the master and slave state
vectors, respectively. A partially linear system is a set of
ordinary differential equations, whose state vector can be divided
into two parts $(\mathbf{u},z)$ in such a way. The two partially
linear systems are coupled through $z$: the $z$ in the slave
system will be the $z$ of the master system. The above coupled
system is said to be projective synchronous if for an initial
condition there is a constant $\beta$ such that asymptotically in
time
\begin{equation}
\|\beta \mathbf{u}_m-\mathbf{u}_s\|\rightarrow 0.
\end{equation}

In this paper, we study the projective synchronization in coupled
fractional order partially linear chaotic systems of the form
\begin{equation}
\begin{array}{l}
\frac{d^\alpha\mathbf{u}_m}{d t^\alpha}=\mathbf{M}(z)\cdot{\mathbf{u}}_m,\\
\frac{d^\alpha z}{d t^\alpha}=f(\mathbf{u}_m,z),\\
\frac{d^\alpha \mathbf{u}_s}{d
t^\alpha}=\mathbf{M}(z)\cdot\mathbf{u}_s
\end{array}
\end{equation}
where $\alpha$ is the fractional order.

In the literature, there are several different definitions of
fractional derivatives, see e.g. [1]. Perhaps the best known one
is the Riemann-Liouville definition:
\begin{equation}
\frac{d^\alpha f(t)}{dt^\alpha} = \frac{1} {\Gamma(n-\alpha)}
\frac{d^n}{dt^n} \int_0^t
\frac{f(\tau)}{(t-\tau)^{\alpha-n+1}}d\tau,
\end{equation}
where $\Gamma(\cdot)$ is the gamma function and $n-1\leq \alpha
<n$. The geometric and physical interpretation of the fractional
derivatives were given in [27]. Upon considering the initial
values to be zero, the Laplace transform of the Riemann-Liouville
fractional derivative is $ L\left\{\frac{d^\alpha
f(t)}{dt^\alpha}\right\}(s)=s^\alpha L\{f(t)\}$. So, the
fractional integral operator of order ``$\alpha$" can be
represented by the transfer function $F(s)=\frac{1}{s^\alpha}$.

According to the standard definition of the fractional
differintegral, we can't directly implement the fractional
operators in time-domain simulations. An efficient method to
circumvent this problem is to approximate (in the frequency
domain) the fractional operators by using the standard integer
order operators. In the following simulations, we will use the
approximation method proposed in [28], which was also adopted in
[9, 11, 12, 14, 15, 16, 17]. In Table 1 of [9], the authors
presented the approximations for $1/s^q$ with $q=0.1-0.9$ in steps
0.1 with errors of approximately 2dB. We will use these
approximations in our following simulations.

An often studied example of the partially linear chaotic system in
projective synchronization in the integer-order case is the Lorenz
system, but, unfortunately the results about the fractional order
Lorenz system is not correct [13], so we cannot use this system as
the example of fractional order partially linear chaotic system in
our study. Here we use the fractional order Chen system [17],
\begin{equation}
\begin{array}{l}
\frac{d^\alpha x}{dt^\alpha}=a(y-x),\\
\frac{d^\alpha y}{dt^\alpha}=(c-a)x-xz+cy,\\
\frac{d^\alpha z}{dt^\alpha}=xy-bz
\end{array}
\end{equation}
which is partially linear with $\mathbf{u}=(x,y)$ and
\begin{equation*}
\mathbf{M}(z)=\left[\begin{array}{cc}-a&a\\c-a-z&c\end{array}\right].
\end{equation*}
Considering current knowledge on fractional order systems, it is
difficult, if not impossible, to analyze projective
synchronization in fractional order systems theoretically. We
numerically study this topic here.

We consider a coupled Chen system with the fractional order
$\alpha=0.9$ in Eq. (3). We let the parameters $(a,b,c)=(35,3,28)$
in Eq. (5), so that the fractional order Chen system is chaotic
[17]. The initial values for the state variables are some random
values close to the origin. The two chaotic oscillators can always
achieve projective synchronization in our simulations by using the
method mentioned above. The dynamical behaviors of the master and
slave systems in a simulation are shown in Fig. 1. In this figure,
the initial values of the coupled system are $(x_m, y_m,z,x_s,y_s)
=(0.0440, 0.0701,0.0610, 0.0300,0.0856)$. In Fig. 1 (a), we show
the time evolution of the scaling factor $x_s/x_m$, which
indicates that the scaling factor converges to a constant value.
In Fig. 1 (b), we show the projections of the master system and
the slave system onto the $x-y$ plane.  As we see that the two
attractors are the same in structure but different in size, which
also clearly indicates the projective synchronization of the
coupled fractional order system.
\begin{figure}[htb]
\centering
\includegraphics[width=14cm]{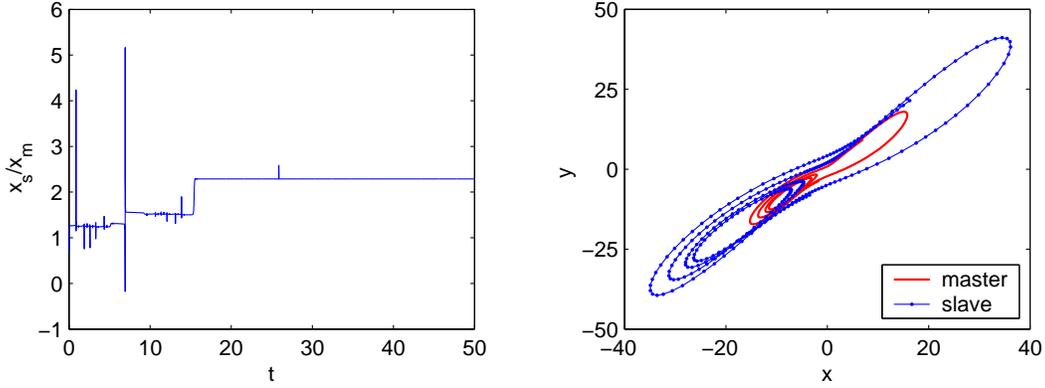}
\caption{(a) Time evolution of the scaling factor $x_s/x_m$. (b)
The projection of the master and slave system onto the $x-y$
plane.}
\end{figure}

Because the scaling factor in projective synchronization is
dependent on the initial condition and unpredictable, in [29, 30]
the authors proposed some control methods for controlling the
scaling factor $\beta$ onto a desired value $\beta^*$. In this
paper, we also present a simple feedback control mechanism for the
coupled fractional order system. We introduce a controller to the
slave system, and the coupled system now becomes
\begin{equation}
\begin{array}{l}
\frac{d^\alpha\mathbf{u}_m}{d t^\alpha}=\mathbf{M}(z)\cdot{\mathbf{u}}_m,\\
\frac{d^\alpha z}{d t^\alpha}=f(\mathbf{u}_m,z),\\
\frac{d^\alpha \mathbf{u}_s}{d
t^\alpha}=\mathbf{M}(z)\cdot\mathbf{u}_s+k(\beta^*\mathbf{u}_m-\mathbf{u}_s)
\end{array}
\end{equation}
where $k$ is a positive constant and $\beta^*$ is the desired
scaling factor. The control mechanism is standard in integer-order
systems (see e.g. [31]), and can be easily understood. If any
component of $\mathbf{u}_s$ is larger than the corresponding
component of $\beta^*\mathbf{u}_m$, then the corresponding
component of $k(\beta^*\mathbf{u}_m-\mathbf{u}_s)$ will be
negative and the rate of that component of $\mathbf{u}_s$ will be
decreased, thus $\mathbf{u}_s/\mathbf{u}_m$ will also be
decreased, and vice versa. So eventually the scaling factor should
asymptotically converge to $\beta^*$.

We let $k=0.1$, and apply the control mechanism to the coupled
system. The desired scaling factor $\beta^*$ can be reached for
any reasonable positive and negative values with any random
initial conditions. In Fig. 2 (a) and (b), we show the projections
of the master and slave systems onto the $x-y$ plane for
$\beta^*=5$ and $-3$, respectively. As we see that the two
attractors in each panel of this figure are the same in structure
but different in size (and direction). The size of the master
system does not change, but the slave system is amplified. By
numerous simulations, we found that, even for a very small $k$
(say $k=0.01$), after a long time, the two oscillators can also be
synchronized up to the constant scaling factor $\beta^*$.

\begin{figure}[htb]
\centering
\includegraphics[width=15cm]{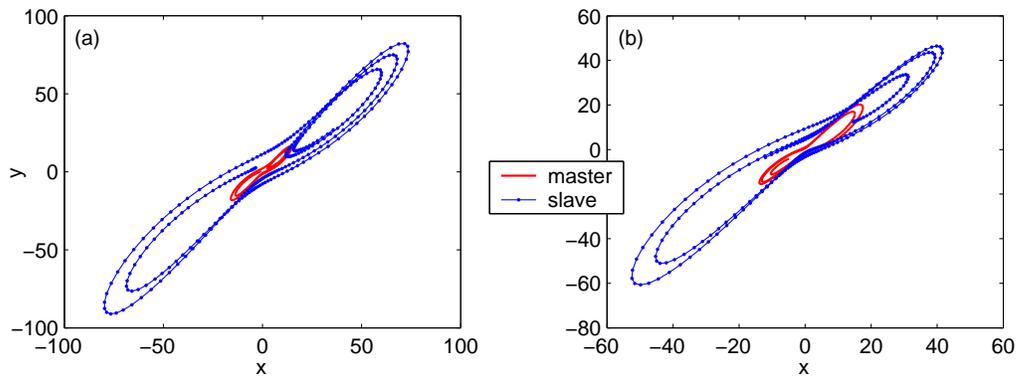}
\caption{The projections of the master and slave systems onto the
$x-y$ plane for (a) $\beta^*=5$; (b) $\beta^*=-3$.}
\end{figure}

For other fractional orders and some other parameter values, we
also found similar phenomena. We omit these results here.

In summary, we have studied the projective synchronization in
coupled two fractional order chaotic oscillators. It is shown that
projective synchronization can also exist in coupled fractional
order chaotic systems. A simple feedback control method is also
presented in this paper, which can drive the scaling factor onto a
desired value. Theoretical analysis of various synchronization
phenomena, including projective synchronization, in fractional
order chaotic systems will be the subject of future research.

This work was supported by the National Natural Science Foundation
of China under Grant 60502009, and the Program for New Century
Excellent Talents in University of China.


\begin{thebibliography}{20}
\bibitem{"[1]"}
I. Podlubny, {\it Fractional Differential Equations}, (Academic
Press, New York, 1999).
\bibitem{"[2]"}
R. Hilfer (Ed.), {\it Applications of Fractional Calculus in
Physics}, (World Scientific, New Jersey, 2001).
\bibitem{"[3]"}
R.L. Bagley, and R.A. Calico, J. Guid., Contr. Dyn. 14, 304
(1991).
\bibitem{"[4]"}
R. C. Koeller, J. Appl. Mech. 51, 299 (1984).
\bibitem{"[5]"}
R.C. Koeller, Acta Mechanica 58, 251 (1986).
\bibitem{"[6]"}
H.H. Sun, A.A. Abdelwahad, and B. Onaral, IEEE Trans. Auto. Contr.
29, 441 (1984).
\bibitem{"[7]"}
M. Ichise, Y. Nagayanagi, and T. Kojima, J. Electroanal. Chem. 33,
253 (1971).
\bibitem{"[8]"}
O. Heaviside, {\it Electromagnetic Theory}, (Chelsea, New York,
1971).
\bibitem{"[9]"}
T.T. Hartley, C.F. Lorenzo, and H. K. Qammer, IEEE Trans. CAS-I
42, 485 (1995).
\bibitem{"[10]"}
P. Arena, R. Caponetto, L. Fortuna, and D. Porto, Proc. ECCTD,
Budapest, 1259 (1997).
\bibitem{"[11]"}
W.M. Ahmad, and J.C. Sprott, Chaos, Solitons and Fractals 16, 339
(2003).
\bibitem{"[12]"}
W.M. Ahmad, and W.M. Harb, Chaos, Solitons and Fractals 18, 693
(2003).
\bibitem{"[13]"}
I. Grigorenko, and E. Grigorenko, Phys. Rev. Lett. 91, 034101
(2003). But the Eq. (5) in this paper is not correct, so the
results presented in this paper are not reliable, which I have
pointed out to the first author of this paper via personal
communication.
\bibitem{"[14]"}
P. Arena, R. Caponetto, L. Fortuna, and D. Porto, Int. J. Bifur.
Chaos 7, 1527 (1998).
\bibitem{"[15]"}
P. Arena, L. Fortuna, and D. Porto, Phys. Rev. E 61, 776 (2000).
\bibitem{"[16]"}
C. Li, G. Chen, Physica A 341, 55 (2004).
\bibitem{"[17]"}
C. Li, G. Chen, Chaos, Solitons and Fractals 22, 549 (2004).
\bibitem{"[18]"}
G.M. Zaslavsky, Phys. Rep. 371, 461 (2002).
\bibitem{"[19]"}
S. Boccaletti, J. Kurths, G. Osipov, D.L. Valladares, and C.S.
Zhou, Phys. Rep. 366, 1 (2002).
\bibitem{"[20]"}
L.M. Pecora and T.L. Carroll, Phys. Rev. Lett. 64, 821 (1990).
\bibitem{"[21]"}
M. Rosenblum, A. Pikovsky, and J. Kurtz, Phys. Rev. Lett. 76, 1804
(1996);\\ A. Pikovsky, M. Rosenlum, G. Osipov, and J. Kurtz,
Physica D 104, 219 (1997).
\bibitem{"[22]"}
M. G. Rosenblum, A.S. Pikovsky, J. Kurths, Phys. Rev. Lett. 78,
4193 (1997).
\bibitem{"[23]"}
R. Mainieri and J. Rehacek, Phys. Rev. Lett. 82, 3042 (1999).
\bibitem{"[24]"}
C. Li, X. Liao, J. Yu, Phys. Rev. E 68, 067203 (2003).
\bibitem{"[25]"}
W.H. Deng, C.P. Li, J. Phys. Soc. Jpn 74, 1045, 2005; \\
J. G. Lu, Chaos, Solitons and Fractals 26, 1125, 2005; \\X. Gao,
Chaos, Solitons and Fractals 26, 141, 2005; \\J. G. Lu, Chaos,
Solitons and Fractals 27, 519, 2006;\\ H. Zhang, C.G. Li, G. Chen,
Int. J. Mod. Phys. C 16, 815, 2005;\\ W.H. Deng, C.P. Li, Physica
A 353, 61, 2005.
\bibitem{"[26]"}
C. Li, Phase and lag synchronization in coupled fractional order
chaotic oscillators, IEEE Trans. Circuits and Systems-II,
submitted.
\bibitem{"[27]"}
M. Moshrefi-Torbati, J.K. Hammond, J. Franklin Inst. 335B, 1077
(1998);\\I. Podlubny, arXiv: math.CA/0110241 (2001).
\bibitem{"[28]"}
A. Charef, H.H. Sun, Y.Y. Tsao, and B. Onaral, IEEE Trans. Auto.
Contr. 37, 1465 (1992).
\bibitem{"[29]"}
D. Xu, Phys. Rev. E 63, 027201 (2001).
\bibitem{"[30]"}
D. Xu, Z. Li, Int. J. Bifur. Chaos 12, 1395 (2002).
\bibitem{"[31]"}
J C. Doyle, B. A. Francis, A. R. Tannenbaum, {\it Feedback Control
Theory} (Macmillan Coll Div, 1992).
\end{thebibliography}
\end{document}